\documentclass[journal]{IEEEtran}
\usepackage{amsmath,amsfonts}
\usepackage[ruled,linesnumbered, longend, vlined]{algorithm2e}
\usepackage{array}
\usepackage{arydshln}
\usepackage{amsthm}
\usepackage[caption=false,font=normalsize,labelfont=sf,textfont=sf]{subfig}
\usepackage{textcomp}
\usepackage{stfloats}
\usepackage{breakurl}
\usepackage{url}
\usepackage{verbatim}
\usepackage{graphicx}
\usepackage{cite}
\usepackage{url}
\usepackage{float}
\usepackage{bm}
\usepackage{multirow}
\usepackage{threeparttable}
\usepackage{hyperref}
\usepackage{amssymb}
\usepackage{booktabs} 
\usepackage[inline]{enumitem}
\newtheorem{remark}{Remark}[section]

\begin{document}

\title{Load Data Valuation in Multi-Energy Systems:\\ An End-to-End Approach}

\author{Yangze Zhou, Qingsong Wen, Jie Song, Xueyuan Cui, and Yi Wang

\thanks{Yangze Zhou, Xueyuan Cui, and Yi Wang are with the Department of Electrical and Electronic Engineering, The University of Hong Kong, Hong Kong SAR, China (e-mail: yzzhou@connect.hku.hk, xycui@eee.hku.hk, yiwang@eee.hku.hk).}
\thanks{Qingsong Wen is with the DAMO Academy, Alibaba Group (U.S.) Inc., Bellevue, WA 98004, USA (e-mail: qingsong.wen@alibaba-inc.com).}
\thanks{Jie Song is with the Department of Industrial Engineering and Management, College of Engineering, Peking University, Beijing, China (email: jie.song@pku.edu.cn).}
}

% The paper headers
\markboth{Submitted to IEEE Transactions on Smart Grid}%
{Shell \MakeLowercase{\textit{et al.}}: A Sample Article Using IEEEtran.cls for IEEE Journals}

\maketitle

\begin{abstract}
Accurate load forecasting serves as the foundation for the flexible operation of multi-energy systems (MES). Multi-energy loads are tightly coupled and exhibit significant uncertainties. Many works focus on enhancing forecasting accuracy by leveraging cross-sector information. However, data owners may not be motivated to share their data unless it leads to substantial benefits. Ensuring a reasonable data valuation can encourage them to share their data willingly. This paper presents an end-to-end framework to quantify multi-energy load data value by integrating forecasting and decision processes. To address optimization problems with integer variables, a two-stage end-to-end model solution is proposed. Moreover, a profit allocation strategy based on contribution to cost savings is investigated to encourage data sharing in MES. The experimental results demonstrate a significant decrease in operation costs, suggesting that the proposed valuation approach more effectively extracts the inherent data value than traditional methods. According to the proposed incentive mechanism, all sectors can benefit from data sharing by improving forecasting accuracy or receiving economic compensation.
\end{abstract}

\section{Introduction}
%subsection{Backgrounds and Motivations}
\IEEEPARstart{M}{ulti}-energy systems (MES) are considered as a promising approach for enhancing energy efficiency and promoting renewable energy accommodation \cite{zhu2022review}. MES coordinates different energy sectors, such as power, cooling, heat, and gas, to satisfy multi-energy loads with high overall energy efficiency \cite{zhou2023can}. Multi-energy load forecasting is the basis for the operations of various energy converters within the MES. For example, in northern China, how much power is generated by combined heat power (CHP) is jointly determined by heat and electricity loads in this area \cite{kang2013balance}. 

Since multi-energy loads are deeply coupled, some studies have utilized cross-sector information to improve load forecasting accuracy. In \cite{li2022multi}, historical load features of different sectors were selected as inputs of the designed parallel architecture. Then, transfer learning is adopted to cope with the issue of data deficiency in MES. \cite{xuan2021multi} introduced multi-task learning with homoscedastic uncertainty to consider the coupling relations, simultaneously achieving better prediction of different sectors. These results all point to an essential fact: data sharing between different sectors in MES is beneficial for forecasting accuracy enhancement.

However, power, gas, and heat/cooling load data are probably owned by different system operators separately. These data owners tend to prioritize their own economic benefits over social benefits when making decisions \cite{cao2017data}. They are generally not motivated to share data, as doing so by sharing cross-sector information does not usually result in substantial advantages for themselves. Only by reasonably valuing the data and fairly allocating the additional profits from data sharing among sectors in MES will they be willing to share their data assets.

Current research about data valuation in energy systems can be primarily divided into two categories but is very limited. The first is to analyze the influence of data on forecasting accuracy. A regression market framework was proposed in \cite{pinson2022regression} to support agents sharing their features with the central agent to train a better regression/forecasting model. The support agents can obtain monetary rewards based on their contribution to improving accuracy. A framework for day-ahead load forecast trading and valuation in an ensemble model was proposed in  \cite{sun2023trading}. Nevertheless, the value of data assets is closely related to the application scenarios and can vary significantly \cite{wang2023data}.  Thus, the second is to study how forecasting accuracy improvements affect operation costs, leading to the quantification of the economic value of data \cite{wang2021data}. \cite{yu2022pricing} utilized Shannon entropy and non-noise ratio as two metrics to measure the quality of photovoltaics-related data and then constructed a reflection from selected metrics to the load forecasting accuracy. On this basis, the economic value of data assets was quantified by simulating a stochastic unit commitment using forecasting results of various accuracy. \cite{goncalves2020towards} proposed a data market based on the electricity market with dual price imbalances to measure the value of the renewable energy data from different owners. 

Although the economic value of data can be quantified by solving an optimization problem to minimize operation costs, the approach of forecasting-then-optimization (FTO) handles forecasting and decision-making as two separate processes. This separated approach may not accurately capture the actual value of data. In FTO, the forecasting model is trained with traditional loss functions such as mean square errors (MSE). These loss functions treat the positive prediction errors and negative prediction errors equally using a quadratic function, whereas they exert different impacts on costs \cite{han2021task}. Specifically, under-forecasts may result in purchases of expensive peaking power while over-forecasting leads to extra generation capacity to be committed \cite{kebriaei2011short}. This fact demonstrates that, in power systems, higher accuracy does not necessarily imply lower costs and there is a discrepancy between minimizing forecasting errors and operation costs \cite{zhang2022cost}. Such discrepancy is more apparent in MES, making it more challenging for the FTO approach to precisely quantify the economic data value. Hence, a more rational approach is to quantify the data value in an end-to-end approach by integrating forecasting and optimization into a single model.

To fill the above research gap, this paper proposes an end-to-end data valuation method within MES. End-to-end refers to combining forecasting and decision-making as a whole. \cite{zhao2021cost} introduces a bilevel programming model that improves both forecasting quality and decision performance by utilizing cost-oriented prediction intervals. \cite{donti2017task} proposed an end-to-end approach in probabilistic machine learning for downstream tasks such as electrical grid scheduling and energy storage arbitrage. End-to-end modeling is an area that has received relatively little attention, let alone its application in value valuation. This paper makes the following contributions:
\begin{enumerate}
    \item Study multi-energy load data valuation problem in MES which has been rarely touched. Specifically, an end-to-end framework is established to quantify the data value from a cost-oriented perspective so that various sectors in MES can collaborate with each other and achieve cost savings. Compared to existing valuation methods, the proposed framework can fully uncover the data value. 
     %  \item Propose a forecasting-optimization integrated framework to study load forecasting and the operation of the MES. This framework combines multiple energy load forecasting and systems operations into one model to achieve cost-oriented load forecasting. 
    \item Develop a two-stage end-to-end model solution to train the forecasting and decision-making integrated end-to-end model, where the decision-making model is not limited to linear programming or quadratic programming problems but can incorporate integer variables. In this way, the method is applicable to various scenarios with integer variables in MES, such as (dis)charging of energy storage and piecewise linear approximation of nonlinear constraints. 
    \item Proposes an intensive mechanism to allocate additional profits from data sharing to different sectors in MES according to their contributions. According to the experiment result, all sectors can benefit from data sharing by either improving forecasting accuracy or receiving economic compensation. 
\end{enumerate}

%\subsection{Paper Organization}
The rest of the paper is structured as follows. Section \ref{PS} defines the problem of end-to-end data valuation within MES. Section \ref{PR} given the preliminaries of this work. Section \ref{e2e} elaborates on the end-to-end modeling to integrate forecasting and optimization. Section \ref{DV} illustrates the details of the proposed data valuation framework via the end-to-end approach. Section \ref{CS} analyzes and visualizes the experiment results. Conclusions and future works can be reached in Section \ref{conclusion}.

\section{Problem Statement}
\label{PS}

\begin{figure}[t]
    \includegraphics[scale=0.75]{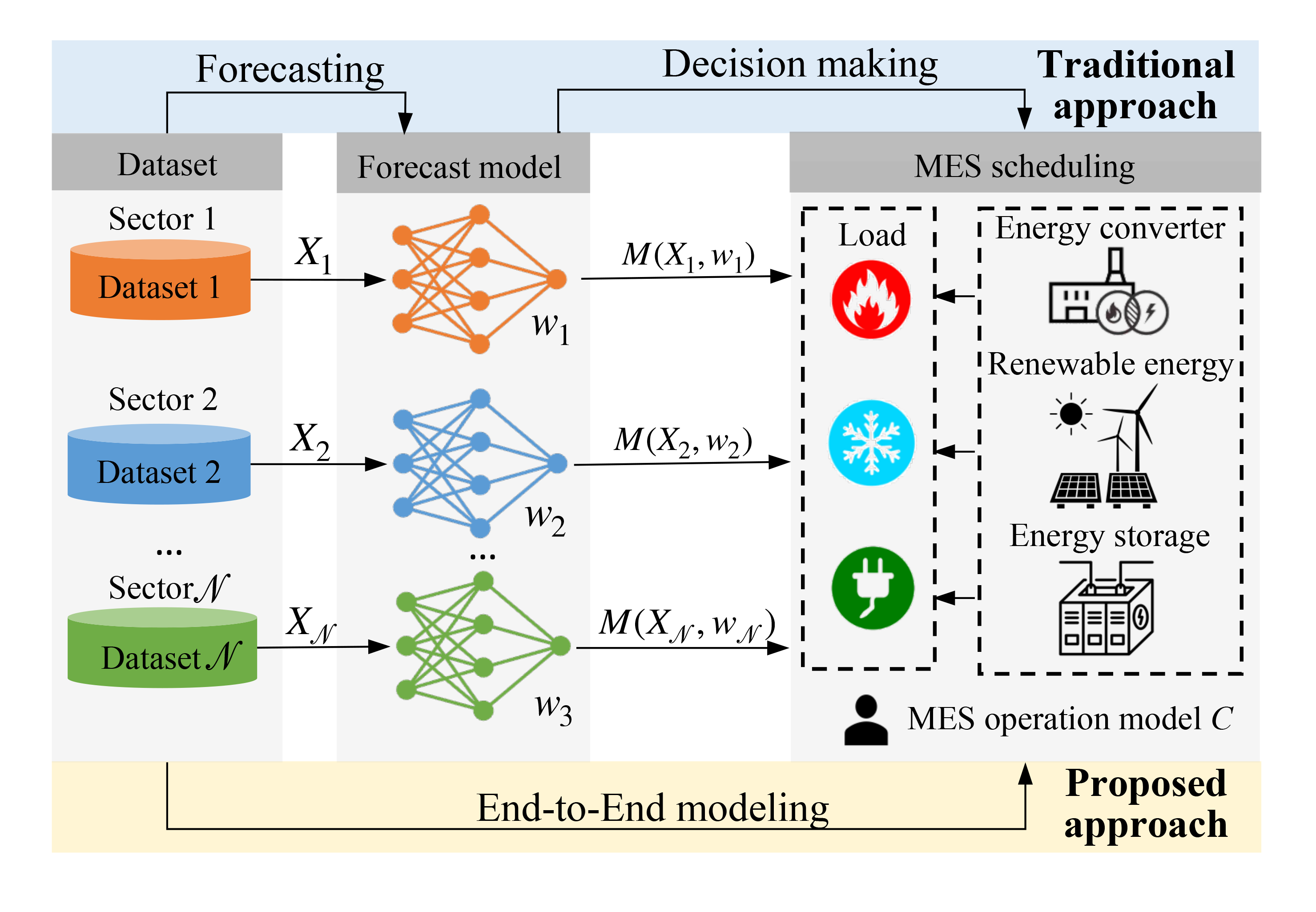}
    \caption{Sketch of the studied scenario. The blue part shows the traditional approach, where the forecasting process and decision-making process are dealt with sequentially. The yellow part is the proposed end-to-end approach.}
    \label{ps}
\end{figure}

We are studying the following scenario shown in Fig. \ref{ps}: In MES with $|\mathcal{N}|$ sectors, an operator is responsible for the scheduling of all equipment within MES based on multi-energy load forecasts $M_n(X_n,w_n)|_{n \in \mathcal{N}}$, submitted by sectors individually, where $X_n$ and $w_n$ denote the input feature and parameters of sector $n$'s forecasting model $M_n$. In our study, we assume that all sectors will report their forecast results honestly, which is commonly satisfied in real-world MES systems. Mathematically, it can be presented as:
\begin{equation}
\label{eq1}
   \mathop{\min}_{z}~ C(z, M_n(X_n,w_n)|_{n \in \mathcal{N}})
\end{equation}
where $C$ and $z$ are the cost and decision variables for the scheduling of MES, respectively.

%\begin{assumptions}
%     This work 
%\end{assumptions}

Traditionally, $w_n$ and $z$ in \eqref{eq1} are determined sequentially, i.e., first training forecasting models by sectors and then scheduling MES by the operator. This practice has two potential limitations:
\begin{enumerate}
    \item Cross-sector data/information has not been shared and fully utilized to enhance forecasting accuracy or reduce operation costs;
    \item The forecasting and decision-making processes are treated separately so that data cannot directly serve final decision-making in MES.  
\end{enumerate}

To overcome the above limitations, the operator can integrate forecasting model training and MES scheduling to form an end-to-end model:  %. The cross-sector information can be shared by the guidance offered by the decision-making process.
\begin{equation}
   \mathop{\min}\limits_{z;w_n|_{n \in \mathcal{N}}} C(z, M_n(X_n,w_n)|_{n \in \mathcal{N}})
\end{equation}
where $w_n$ and $z$ are optimized as a whole; the operation costs is denoted as $C_{\mathcal{N}}$ if all sectors cooperate with the MES operator. The cooperation means the sectors share their data $X_n$ with the operator indirectly.

To encourage sectors to participate in the end-to-end model, the value of the data owned by various sectors should be quantified. In this work, the data value is defined as the additional profits derived from data sharing in the cooperation between sectors and MES operators. Hence, two essential problems need to be solved:
\begin{enumerate}
    \item How many additional profits $V(\mathcal{N})$ can be derived from data sharing of various sectors in MES \text{?}
    \item How to make a fair plan $\{v_1,v_2,\cdots,v_\mathcal{N}\}$ to allocate the profits $V(\mathcal{N})$ to each sector \text{?}
\end{enumerate}

\section{Preliminaries}
\label{PR}
This section provides preliminaries for the two basic components of data valuations: the MES load forecasting model $M_n$ and the operation model $C$. We choose LSTM and the model proposed in \cite{huang2020matrix} for $M_n$ and $C$, respectively.
Note that $M_n$ can be any forecast model that can be trained by gradient descents and  $C$ can be any mixed integer linear programming (MILP) problem. 

\subsection{Forecasting Module $M_n$}
\begin{figure}[t]
    \centering
    \includegraphics[scale=0.45]{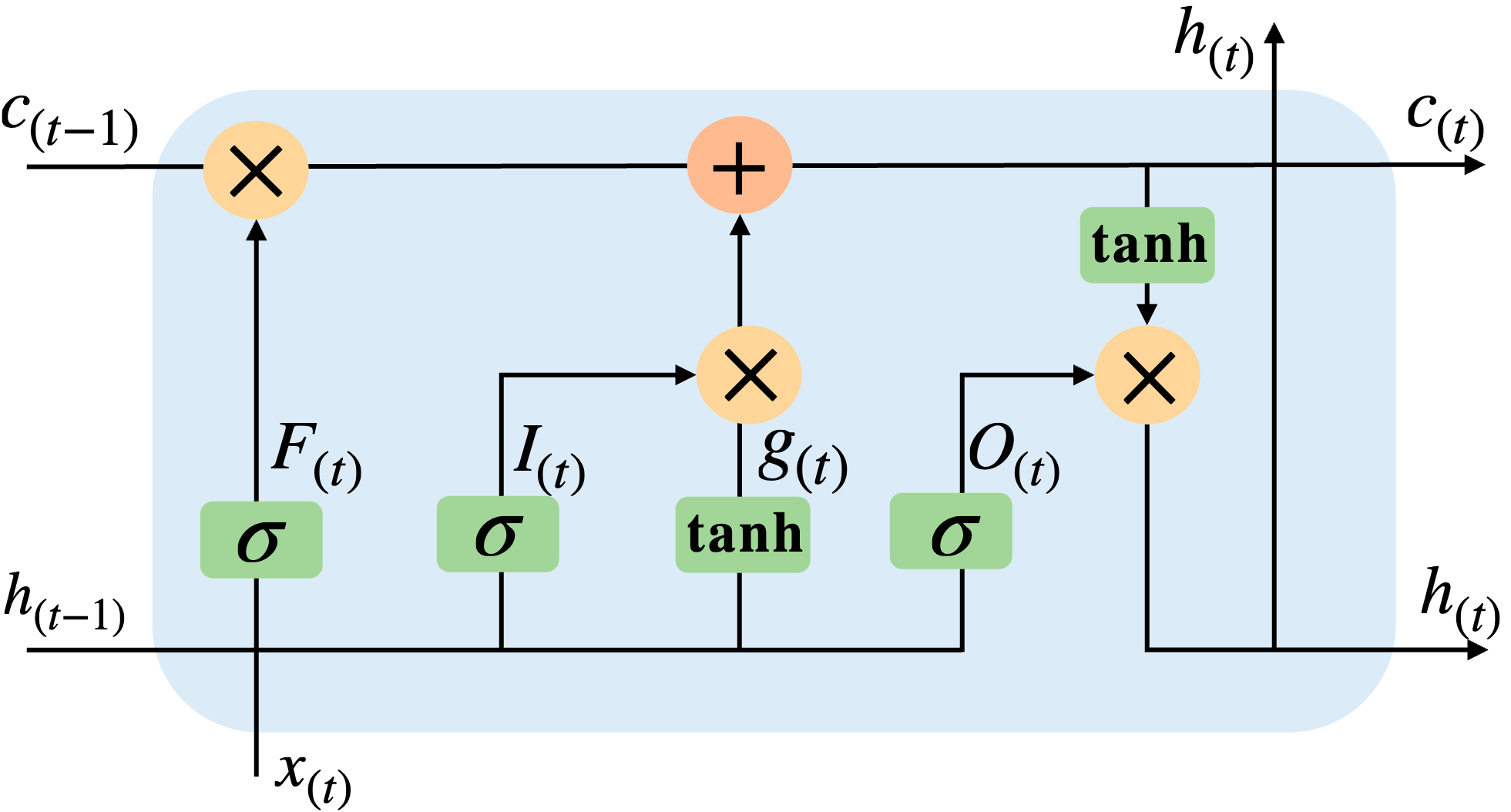}
    \caption{Sketch of an LSTM unit}
    \label{lstm}
\end{figure}

LSTM is a recurrent neural network with several kinds of special gates to capture long-term and temporal dependence\cite{yu2019review}. It has been widely applied for forecasting electricity load \cite{kong2017short}, heat load \cite{chung2022district}, and cooling load \cite{zhang2020hybrid}. An LSTM unit is shown in Fig. \ref{lstm}, and its working principles can be formulated as follows:
\begin{subequations}
\begin{equation}{{\mathcal{F}}{\left( t \right)}}=\sigma\left( W_{x,f}^{T}{{x}{\left( t \right)}}+W_{h,f}^{T}{{h}{\left( t-1 \right)}}+{{b}_{f}} \right)
     \label{LSTM_principle_1}\end{equation}
\begin{equation}{{\mathcal{I}}{\left( t \right)}}=\sigma\left( W_{x,i}^{T}{{x}{\left( t \right)}}+W_{h,i}^{T}{{h}{\left( t-1 \right)}}+{{b}_{i}} \right)\label{LSTM_principle_2}\end{equation}
\begin{equation}{{\mathcal{O}}{\left( t \right)}}=\sigma\left( W_{x,o}^{T}{{x}{\left( t \right)}}+W_{h,o}^{T}{{h}{\left( t-1 \right)}}+{{b}_{o}} \right)
     \label{LSTM_principle_3}\end{equation}
\begin{equation}{{{g}}{\left( t \right)}}=tanh\left( W_{x,g}^{T}{{x}{\left( t \right)}}+W_{h,g}^{T}{{h}{\left( t-1 \right)}}+{{b}_{g}} \right)
     \label{LSTM_principle_4}\end{equation}
\begin{equation}{{c}{\left( t \right)}}={{\mathcal{F}}{\left( t \right)}}\otimes {{c}{\left( t-1 \right)}}+{{\mathcal{I}}{\left( t \right)}}\otimes {{g}{\left( t \right)}}
     \label{LSTM_principle_5}\end{equation}
\begin{equation}{{y}{\left( t \right)}}={{h}{\left( t \right)}}={{\mathcal{O}}{\left( t \right)}}\otimes tanh\left( {{c}{\left( t \right)}} \right)
     \label{LSTM_principle_6}\end{equation}
\end{subequations}
where $x{(t)}$, $g{(t)}$, $y{(t)}$ are inputs features, intermediate state, and output at $t$, $h{(t-1)}$ and $c{(t-1)}$ are hidden state and memory cell propagated from the $t-1$, $W_{x,i}$, $W_{x,f}$, $W_{x,o}$, $W_{x,g}$ are weights related to $x$, $W_{h,i}$, $W_{h,f}$, $W_{h,o}$, $W_{h,g}$ are weights releated to hidden state $h$, and $b_{i}, b_{f}, b_{o}, b_{g}$ are corresponding bias terms, $\sigma$ and $tanh$ are sigmoid and tanh activation function, $\otimes$ denotes the Hadamard product. 

\subsection{Decision Making Module $C$}
This work develops the model of MES with the concept of Energy Hub (EH) \cite{wang2018mixed}. EH is defined as a unit that provides the functions of input, output, conversion, and storage of multiple sectors \cite{mohammadi2017energy}. The comprehensive energy flow equations of the EH are formulated as \eqref{EH} \cite{huang2020matrix}.
\begin{equation}
\label{EH}
[\textbf{X}^{T},\textbf{Y}^{T},\textbf{Z}^{T}]\textbf{V}=[\textbf{V}_{in}^{T},\textbf{V}_{out}^{T},\textbf{0}^{T}]^{T}
\end{equation}
where $\textbf{V}_{in}$ is input vector, $\textbf{V}_{out}$ is output vector, $\textbf{V}$ represent energy flows in all branches, $\textbf{X}$ is input incidence matrix, $\textbf{Y}$ is output incidence matrix, matrix $\textbf{Z}$ combines the nodal energy conversion matrix of all nodes in
the EH.

In this work, variable converter efficiency is considered, which introduces non-linear constraints into EH modeling due to the variable multiplication. The piecewise linear approximation has been applied to depict non-linear input-output relationships in a linear way \cite{huang2020matrix}. Additionally, this work includes the consideration of energy storage equipment.

In the day-ahead schedule strategy stage, the MES operator schedules the equipment in different sectors according to the day-ahead forecasts. The optimization problem of the day-ahead strategy can be formulated as:
\begin{equation}
    \begin{aligned}
   % &\mathop{\arg\min}\limits_{M \in \vartheta} c(M,\hat{L})=\sum_{t=1}^{t=T} \mathbf{F}_{a}\cdot I_{a}\\
    \min &~~\sum_{t=1}^{t_{24}} \textbf{F}_{t} \cdot \textbf{V}_{in,t}^{T} \quad \text{s.t.} \quad \eqref{EH}
    \end{aligned}
    \label{MES_opt}
\end{equation}
    
where $\textbf{V}_{in,t}$ and $\textbf{F}_{t}$ denote the vector of input and corresponding prices at time $t$ in the day ahead.

In the intra-day process, the real-time energy loads are likely to differ from the forecasting load generally. The goal of the intra-day schedule strategy is to maintain the energy balance of different sectors via temporary scheduling or energy storage charging/discharging with the lowest cost:
\begin{equation}
\begin{aligned}
     &\min \sum_{t=1}^{t_{24}}(\widetilde{\textbf{F}}_{t}\cdot \widetilde{\textbf{V}}_{in,t}^{T}
    +F_{ch}\cdot Q_{ch,t}+ F_{dis}\cdot Q_{dis,t})\\
    &\text{s.t.} \quad \eqref{EH}\\
     &\textbf{V}_{in,t}-RD \leqslant \widetilde{\textbf{V}}_{in,t}\leqslant \textbf{V}_{in,t}+RU, \\
     & M^{\text{real}}_{t}=\widetilde{\textbf{V}}_{out,t}+Q_{dis,t}-Q_{ch,t}&
\end{aligned}
\end{equation}
where $RD$ and $RU$ are down/up reserve capacity, $\widetilde{\textbf{V}}_{in,t}$, $\widetilde{\textbf{V}}_{out,t}$ and $\widetilde{\textbf{F}}_{t}$ denote the input vector, output vector, and corresponding prices at time $t$ in the intra-day, $M^{\text{real}}_{t}$ are the real-time energy loads at time $t$, $Q_{dis,t}$ and $Q_{ch,t}$ are the discharging power and charging power of the energy storage at time $t$, $F_{dis}$ and $F_{ch}$ are the cost of discharging and charging energy storage.

\section{End to End Optimziaiton}
\label{e2e}
This section details the proposed end-to-end model, including modeling and solution algorithm. 

\begin{figure*}[t]
\centering
\includegraphics[scale=0.75]
    {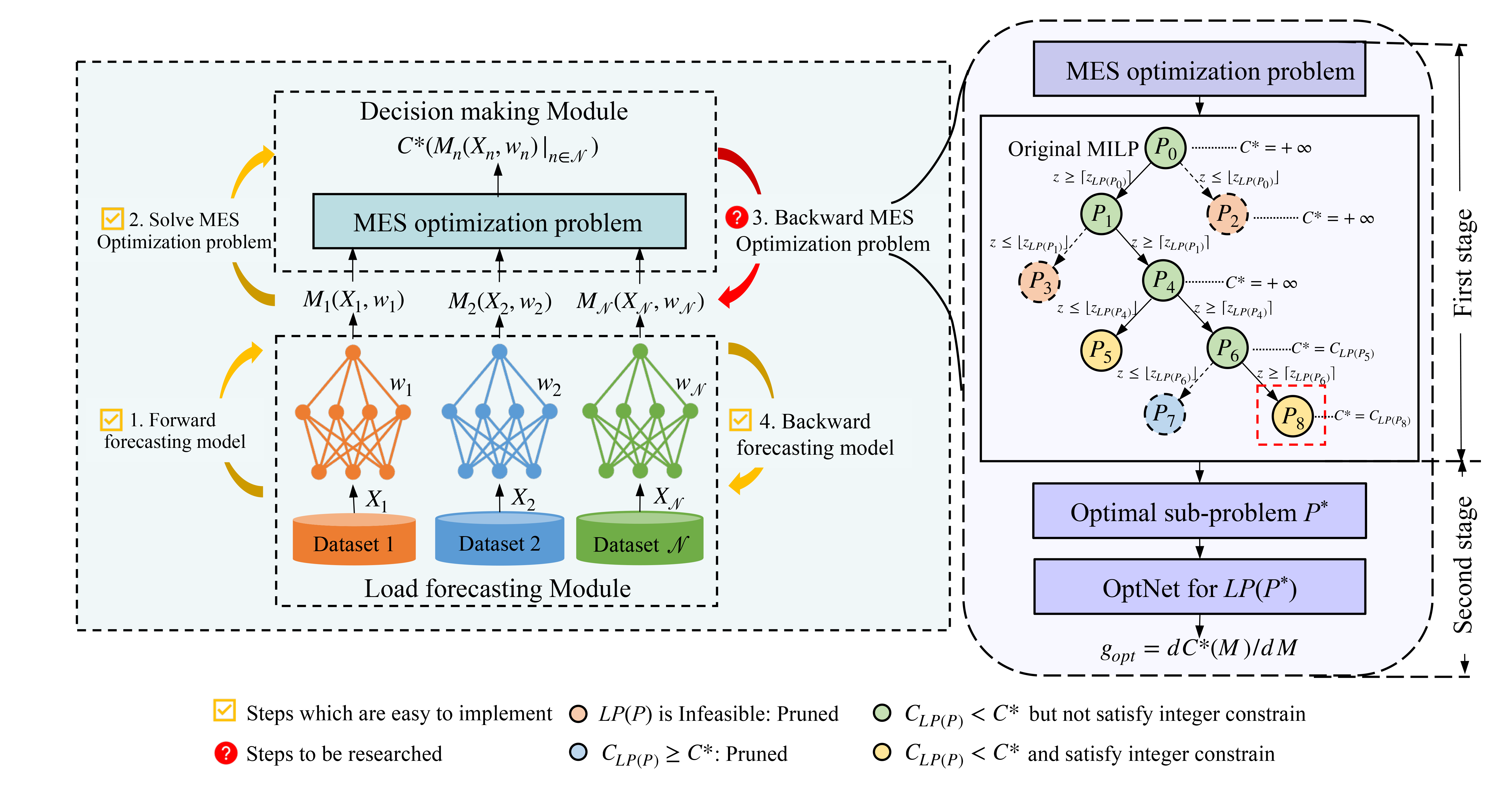}
    \caption{The sketch of end-to-end modeling. The left-hand part is the four main procedures of end-to-end modeling. The right-hand part is the proposed two-stage end-to-end modeling solution algorithm to integrate forecasting models with a MILP problem. }% The yellow arrow represents the steps that can be conducted easily, while the red arrow indicates the main challenge of end-to-end modeling.}
    \label{end-to-end sketch}
\end{figure*}

\subsection{End-to-End Modeling}
End-to-end modeling refers to integrating the training of the local forecasting module $M_n$ and the MES optimization problem $C$ as a whole, as shown in the left-hand part of Fig. \ref{end-to-end sketch}. An intuitive idea to train the end-to-end model is forward and backward propagation, as used for traditional neural network training. Thus, we have four steps for the model training:
\begin{enumerate}
    \item Forward load forecasting models: Each sector inputs the feature $X_n$ to the forecasting model and then outputs load forecasts $M_n(X_n,w_n)$.
    \item Solve MES optimization problem: The operator collects all sectors' forecasts $M=M_n(X_n,w_n)|_{n\in \mathcal{N}}$ as the inputs of the MES optimization problem to obtain the optimal schedule $z^{*}$ and operation costs $C(z^{*}, M_n(X_n,w_n)|_{n\in\mathcal{N}})$.
    \item Backward MES optimization problem: The operator gives back information about the relationship (i.e., gradients) between operation costs and forecasts to the load forecasting models.
    \item Backward load forecasting models: Each sector receives the gradients from the operator and updates its model parameter $w_n$ by backwarding the forecasting model.
\end{enumerate}
Steps 1, 2, and 4, marked by yellow, can be easily conducted as usual, while
the main difficulty is step 3, marked by red: how to obtain the gradient of cost $C$ over load forecasts $M$.

The MES optimization problem \eqref{MES_opt} can be abstracted as follows:
\begin{equation}
    \begin{array}{cl}\label{opt problem}
\operatorname{min} & C(z,M), \\
\text { s.t. } & f(z,M) \leq 0, h(z,M) = 0.
\end{array}
\end{equation}
where $C$, $z$, and $M$ denote the objective function, decision variables, and multi-energy load forecasts, respectively; $f$ and $h$ denote inequality and equality constraints, respectively. 

%where M∈R|N|M \in \mathbb{R}^{|\mathcal{N}|} are muilt-energy laod forecasts; z∈Rdz \in \mathbb{R}^{d} is the decision variable; C:R|N|×Rd↦RC: \mathbb{R}^{|\mathcal{N}|} \times \mathbb{R}^{d} \mapsto \mathbb{R} is the objective function to be optimized, f:R|N|×Rd↦Rpf: \mathbb{R}^{|\mathcal{N}|} \times \mathbb{R}^{d} \mapsto \mathbb{R}^{p} is inequality constraint (pp is the number of inequality constraints), h:R|N|×Rd↦Rqh: \mathbb{R}^{|\mathcal{N}|} \times \mathbb{R}^{d} \mapsto \mathbb{R}^q is equality constraint (qq is the number of equality constraints). 

For given $M$, the optimal operation costs is:
\begin{equation}
    C^{*}(M)=\inf \left\{C(z,M) \mid f(z,M) \leq 0, h(z,M)=0\right\}
\end{equation}
The optimal decision $z^{*}$ is dependent on the forecasts $M$:
\begin{equation}
    z^{*}(M)=\left\{z \mid C(z,M)=C^{*}(M), f(z,M) \leq 0, h(z,M)=0\right\}
\end{equation}

To train the end-to-end model that minimizes the operation costs $C^*(M)$, it is necessary to calculate the gradient of $C^*(M)$ over $M$. For a quadratic programming problem, \cite{2017OptNet} introduced an optimization differentiable neural network (OptNet) to compute the gradient indirectly by solving equations from the KKT condition of a Lagrangian function.

The Lagrangian function of the optimization problem \eqref{opt problem} is:
\begin{equation}
    \mathcal{L}(z,\lambda,\mu,M)=C(z,M)+\lambda^{T}f(z,M)+\mu^{T}h(z,M), 
\end{equation}
 where $\lambda$ and $\mu$ are dual variables.
 The KKT condition of $\mathcal{L}(z,\lambda,\mu,M)$ is:
\begin{equation}
\label{KKT}
    \left\{
                \begin{array}{lllll}
                f(z,M)\leq 0, \\ 
                h(z,M) = 0, \\ 
                \lambda_{i}\geq 0, \quad  i\in {1,2,\dots,q}\\
                \lambda_{i}f_{i}(z,M)=0, \quad  i\in {1,2,\dots,q}\\
                \nabla_z \mathcal{L}(z,\lambda,\mu,M)=0.
                \end{array}
              \right.
\end{equation}
Denoted $[z,\lambda,\mu]$ as $\tilde{z}$, a implicit function 
$G(\tilde{z},M)$ can be derived from \eqref{KKT}:
\begin{equation}
G(\tilde{z},M)=G(z,\lambda,\mu,M)=\left[\begin{array}{c}
        \nabla_z \mathcal{L}(z,\lambda,\mu,M)\\
        \lambda f(z,M)\\
       h(z,M)\\
    \end{array}\right]=0.
\end{equation}
The gradient of $\tilde{z}$ over $M$ can be obtained by
the differential principle of implicit function:
\begin{equation}\label{opt_gradient}
    \frac{d\tilde{z}}{d M}=G_{\tilde{z}}^{-1}(\tilde{z},M)G_{M}(\tilde{z},M),
\end{equation}
where
\begin{equation}
    \frac{d\tilde{z}}{d M}=\left[
    \begin{array}{c}
        \frac {\partial z} {\partial M} \quad
        \frac {\partial \lambda} {\partial M} \quad
        \frac {\partial \mu} {\partial M} \quad 
    \end{array}\right]^{T} \nonumber
\end{equation}
\begin{equation}
\begin{aligned}
&G_{\tilde{z}}(\tilde{z},M)\\&= \left[
    \begin{array}{ccc}
        \frac {\partial \nabla_z \mathcal{L}(z,\lambda,\mu,M)} {\partial z}&        \frac {\partial \nabla_z \mathcal{L}(z,\lambda,\mu,M)} {\partial \lambda}&\frac {\partial \nabla_z \mathcal{L}(z,\lambda,\mu,M)} {\partial \mu}\\
        \frac{\partial \lambda f(z, M)}{\partial z}
    &\frac{\partial \lambda f(z, M)}{\partial \lambda}&\frac{\partial \lambda f(z, M)}{\partial \mu}\\
            \frac {\partial h(z,M)} {\partial z}&\frac {\partial h(z,M)} {\partial \lambda}&\frac {\partial h(z,M)} {\partial \mu}
    \end{array}\right] \\&=
 \left[
    \begin{array}{ccc}
        \frac {\partial \nabla_z \mathcal{L}(z,\lambda,\mu,M)} {\partial z}&(\frac {\partial f(z, M)} {\partial z})^{T}&(\frac {\partial h(z,M)} {\partial z})^{T}\\
            \text{diag}(\lambda)\frac {\partial f(z, M)} {\partial z}&\text{diag}(f(z, M))&0\\
            \frac {\partial h(z,M)} {\partial z}&0&0\\
    \end{array}\right] \\ \nonumber
\end{aligned}
\end{equation}

\begin{equation}
\begin{aligned}
    G_{M}(\tilde{z},M)&= \left[
    \begin{array}{c}
        \frac {\partial \nabla_z \mathcal{L}(z,\lambda,\mu,M)} {\partial z}\\
            \text{diag}(\lambda)\frac{\partial f(z, M)}{\partial z}\\
            \frac{\partial h(z,M)}{\partial z}
    \end{array}\right]
\end{aligned}
\end{equation}

The gradient of $C^{*}(M)$ over $M$ can be divided as follows according to the chain principle:
\begin{equation}\label{gradient}
    g_\text{opt}=\frac{d C^{*}(M)}{d M}=\frac{d C^{*}(M)}{d z}\frac{d z}{d M}.
\end{equation}
where ${d C^{*}(M)}/{d z}$ can be easily calculated based on the explicit function $C(z,M)$; ${d z}/{d M}$ can be computed using \eqref{opt_gradient}. Therefore, the whole forward and backward propagation process for the end-to-end model training in Fig. \ref{end-to-end sketch} can be successfully implemented.

The OptNet was originally designed for linear or quadratic programming problems. However, the MES optimization problem will inevitably introduce integer variables if we want to model battery status, nonlinear efficiencies, etc. \cite{urbanucci2018limits}, which is usually modeled as an MILP problem. OPTNet is not applicable in this case. The next subsection will explain how to solve the end-to-end model when the forecasting models are integrated with an MILP problem.

\subsection{Solution Algorithm} 
When the forecasting models are integrated with an MILP problem, $dC^{*}(M)/dz$ either becomes zero or does not exist \cite{poganvcic2019differentiation}. This makes it challenging to solve the end-to-end modeling by directly utilizing the original OptNet. To address this issue, this study proposes a two-stage end-to-end model solution method, as shown in the right-hand part of Fig. \ref{end-to-end sketch}. Algorithm \ref{alg1} details the proposed method, which includes two main stages: optimal sub-problem selection and optimal sub-problem backward propagation. 

In the first stage, the branch and bound method is utilized to select the optimal sub-problem of the original MILP problem. Firstly, a search tree $T$ is initialized with the original MILP problem $P_0(z,M)$, each node of which is a sub-problem of $P_0(z,M)$; variables $C^{*}$, $z^{*}$, and $P^{*}$, are defined to store the temporary search results as well. Then, we move to the branch and bound phases. In the bound phase, the child nodes are solved to obtain the upper and lower bounds of operation costs \cite{zhao2021cost}. Specifically, a sub-problem $P$ is popped from the search tree $T$ unless $T$ is empty, and its linear relaxation problem $LP(P)$ is solved. If $LP(P)$ is unfeasible or $C_{LP(P)}$ is larger than $C^{*}$, it ought to be pruned; otherwise, the integer constraints should be checked. If the $z_{P}$ satisfies the integer constraints, variables $C^{*}$, $z^{*}$, and $P^{*}$ should be updated. If a certain variable $z_j$ violates the integer constraint, it turns to the branch phase. In the branch phase, the problem $P$ will be divided into child nodes by branching on an integer variable  \cite{fletcher1998numerical}. The feasible space of $z_j$ will be divided into two parts to create two new sub-problems, $P_l$ and $P_r$, which will be added to the search tree $T$. The branch and bound phases will continue until there are no child nodes that have not been fathomed yet. The first stage will return the optimal operation costs $C^{*}(M)$, decision variable $z^{*}$ and sub-problem $P^{*}$ of the original MILP problem $P_0(z,M)$ according to \eqref{opt_gradient}.

In the second stage, the backward propagation is implemented for the optimal sub-problems $LP(P^{*})$. An OptNet is first developed with the $LP(P^{*})$. On this basis, the gradient of $C^{*}(M)$ over the $M$, i.e., $g_\text{opt}$ can be computed by back-propagating through the OptNet.

\textbf{Discussion}: Another intuitive idea to address the issue of non-differentiability is to incorporate OptNet into the branch and bound search process, i.e., the OptNet-embedded branch and bound method. This solution requires a new temporary variable $g^{*}$ to store the temporary gradient calculation result and OptNet is embedded in the bound phase. Specifically, for each child node that satisfies $C_{LP(P)} < C^{*}$ and integer constrain (yellow node in Fig. \ref{end-to-end sketch}), we need to construct an OptNet for $LP(P)$ and backward it to update $g^{*}$ accordingly in line 8 of Algorithm \ref{alg1}. 

\begin{remark}
    The gradient of $dC^{*}(M)$ over $M$ derived by the OptNet-embedded branch and bound solution method is the same as the two-stage solution approach.
\end{remark}
\begin{proof}
The incorporation and backpropagation of OptNet have no impact on the bound and branch processes. As a result, the OptNet-embedded branch and bound method will select the same optimal sub-problem as the first stage of the proposed method. It is evident that the gradient of the objective function for the same sub-problem over $M$ remains unchanged.
\end{proof}
Given that the proposed two-stage end-to-end model solution enables the gradient calculation with less computational complexity and storage requirements, we have opted for it for our end-to-end model training.

\begin{comment}
\SetAlgoVlined
\begin{algorithm}[t]
\caption{Two-stage End-to-End Mdoel Solution}\label{alg1}
\SetKwInOut{KIN}{Input}
\SetKwInOut{KOUT}{Output}
\KIN{Original MILP problem P0(z,M)P_0(z,M)}
\KOUT{The gradient of C∗(M)C^{*}(M) over MM}

\SetKwProg{Fn}{Optimal sub-problem selection}{:}{}
  \Fn{}{
Initialize T:=[P0(z,M)]T:=[P_0(z,M)], C∗:=+∞C^{*}:=+\infty, z∗=∅z^{*}=\varnothing, P∗=∅P^{*}=\varnothing

\While{T≠∅T\neq \varnothing}{
Pop PP from TT\;
\tcp{Bound Phase}
CP,zP=C_P, z_P = \text{Solve} LP(P)LP(P)\;
\If {LP(P)LP(P) is feasible \textbf{and} CP<C∗C_P < C^{*}}{\eIf{zPz_P \text{satisfy the integer constraints}}{
C∗=CPC^{*}=C_P\;
z∗=zPz^{*}=z_P\;
P∗=PP^{*}=P\;
}
{\tcp*[h]{Branch Phase}\\
    Pl=P∧zj≤⌊zP,j⌋P_l=P\land z_{j} \leq \lfloor z_{P,j}\rfloor\;
    Pr=P∧zj≥⌈zP,j⌉P_r=P\land z_{j} \geq \lceil z_{P,j}\rceil\;
    T∪{Pl,Pr}T \cup \{P_l,P_r\}\;
}
}
}

\textbf{Return C∗C^{*}, P∗P^{*}, and z∗z^{*}} 
}

\SetKwProg{Fn}{Optimal sub-problem backward propagation}{:}{}
  \Fn{}{
Construct \text{OptNet} for LP(P∗)LP(P^{*})\;
goptg_{\text{opt}} = \text{Backward}(OptNet)\;
\textbf{Return goptg_{\text{opt}}} \;
}

\end{algorithm}

\end{algorithm}
\end{comment}

\SetAlgoVlined
\begin{algorithm}[t]
\caption{Two-stage End-to-End Mdoel Solution}\label{alg1}
\SetKwInOut{KIN}{Input}
\SetKwInOut{KOUT}{Output}
\KIN{Original MILP problem $P_0(z,M)$}
\KOUT{The gradient of $C^{*}(M)$ over $M$}

\SetKwProg{Fn}{Optimal sub-problem selection}{:}{}
  \Fn{}{
Initialize $T:=[P_0(z,M)]$, $C^{*}:=+\infty$, $z^{*}=\varnothing$, $P^{*}=\varnothing$

\While{$T\neq \varnothing$}{
Pop $P$ from $T$\;
\tcp{Bound Phase}
$C_{LP(P)}, z_{LP(P)} = $\text{Solve} $LP(P)$\;
\If {$LP(P)$ is feasible \textbf{and} $C_{LP(P)} < C^{*}$}{\eIf{$z_P$ \text{satisfy the integer constraints}}{
$C^{*}=C_{LP(P)}$, $z^{*}= z_{LP(P)}$, $P^{*}=P$\;
}
{\tcp*[h]{Branch Phase}\\
    $P_l=P\land z_{j} \leq \lfloor z_{LP(P),j}\rfloor$\;
    $P_r=P\land z_{j} \geq \lceil z_{LP(P),j}\rceil$\;
    $T \cup \{P_l,P_r\}$\;
}
}
}

\textbf{Return $C^{*}$, $P^{*}$, and $z^{*}$} 
}

\SetKwProg{Fn}{Optimal sub-problem backward propagation}{:}{}
  \Fn{}{
Construct \text{OptNet} for $LP(P^{*})$\;
$g_{\text{opt}}$ = \text{Backward}(OptNet)\;
\textbf{Return $g_{\text{opt}}$} \;
}

\end{algorithm}

\section{Data Valuation Framework}
\label{DV}

This section will elucidate the proposed data valuation framework based on the end-to-end model, including additional profit quantification and allocation.

\subsection{Additional Profit Quantification}

The end-to-end model enables various sectors to achieve data indirect data sharing: cross-sector information is first integrated into the optimization problem of the MES schedule; this integrated information is then transmitted back to sectors in MES by the gradient $g_\text{opt}$. Compared to traditional FTO methods, the indirect cross-sector information sharing in the cooperation between sectors and the MES operator helps to decrease operation costs $C^{*}(M)$. The reduced operation costs can be regarded as the additional profits derived from the data sharing. The additional profits $V(\mathcal{N})$ is defined as:
\begin{equation}\label{valuation}
    V(\mathcal{N})=C_{\mathcal{N}}-C_{\varnothing}
\end{equation}
where $C_{\mathcal{N}}$ is the operation costs if all sectors cooperate with the MES operator in the end-to-end model, and $C_{\varnothing}$ is the operation costs of the traditional FTO approach.

%The proposed end-to-end dava valuation includes basic model development and end-to-end data valuation. 
The proposed method for quantifying the additional profits is illustrated in Algorithm \ref{alg2}, which consists of four steps.
\begin{enumerate}
    \item Each sector $n \in \mathcal{N}$ utilizes its own data to develop the basic forecasting model $M_{n}$. This step can be used for the next step calculation and also provides good initial parameters for end-to-end model training.
    \item The operation costs of the traditional FTO approach $C_{\varnothing}$ is computed with the forecasts provided by the basic forecasting model $M_{n}$. 
    \item Operator integrates the forecasting model with the MES optimization problem for end-to-end model training. Firstly, the MES operator receives the forecasts $M$ and selects the optimal sub-problems $P^{*}$ of the original MILP issue $P(z,M)$. Secondly, the MES operator develops the OptNet for the $LP(P^{*})$ and acquires the gradient of $C^{*}$ over $M$ by backward propagating the OptNet. Thirdly, sectors in MES receive the gradient of the operation costs $C^{*}(M)$ over their own forecasts $M_n(X_n,w_n)$, which is denoted as $g_{\text{opt},n}$, and update their forecasting model $w_n$. These three steps will continue until convergences or reach the maximum training epoch. 
    \item Operator forward propagates the end-to-end model to calculates the operation costs $C_\mathcal{N}$ and then quantifies the additional profits using \eqref{valuation}.
\end{enumerate}

After quantifying the additional profits, it is also essential to allocate them fairly among sectors in order to encourage their cooperation with the MES operator.

\begin{algorithm}[t]
\caption{Additioanl Profit Quantification} %算法的名字
\label{alg2}
\LinesNumbered %要求显示行号
\DontPrintSemicolon
\SetKwFunction{Execution}{\textit{Execution}}
\SetKwInOut{KIN}{Input}
\SetKwInOut{KData}{Data}
\SetKwInOut{KOUT}{Output}
\KIN{Sectors set $\mathcal{N}$, Traditional loss function $L_{MSE}$, Maximum basic model
training epoch $E_1$, Maximum end-to-end model training epoch $E_2$, MILP optimization problem $P(\cdot,\cdot)$\;}
\KOUT{Value of data owned by $\mathcal{N}$.}

\SetKwProg{Fn}{Basic model development}{:}{}
  \Fn{}{

\For {each sector $n \in \mathcal{N}$}{
Random initialize parameters $w_n|_{n\in \mathcal{N}}$\;
\For {$k \in [0,E_1]$}{
$M_n=M_n(X_n;w_n^{(k)})$\;
$g_n=\text{Backward}(L_{MSE}(M_n,M^{\text{real}}_n))$\;
$w_n^{(k+1)}=w_n^{(k)}-lr\cdot g_n$\;
}
}
\textbf{Return} $M_{n}|_{n \in \mathcal{N}}$
}

\SetKwProg{Fn}{End-to-End data valuation}{:}{}
  \Fn{}{
\SetKwProg{Fn}{$C_{\varnothing}$ Calculation}{:}{}
  \Fn{}{
$C_{\varnothing}=\mathop{\min}\limits_{z} C(z,M_n(X_n,w_n)|_{n \in \mathcal{N}})$
}

\SetKwProg{Fn}{End-to-End modeling}{:}{}
  \Fn{}{
\For {$k \in [0,E_2]$}{
$M=M_n(X_n,w_n)|_{n \in \mathcal{N}}$\;
$P^{*} =$ Optimal sub-problem of $P(z,M)$\;
Construct \text{OptNet} for $LP(P^{*})$\;
\For {sector $n \in \mathcal{N}$}{
$g_{\text{opt},n}=\text{Backward}(\text{OptNet})$\;
$g_{n}=\text{Backward}(g_{\text{opt},n})$\;
$w_n^{(k+1)}=w_n^{(k)}-lr\cdot g_n$\;
}
}

}
\SetKwProg{Fn}{$C_{\mathcal{N}}$ Calculation}{:}{}
  \Fn{}{
$
    C_{\mathcal{N}}=\mathop{\min}\limits_{z} C(z, M_n(X_n,w_n)|_{n \in \mathcal{N}})
$
}
\SetKwProg{Fn}{Additioanl profits quantification}{:}{}
  \Fn{}{
$ V(\mathcal{N}) = C_{\mathcal{N}}-C_{\varnothing}\;$
}
\textbf{Return} $V(\mathcal{N})$
}
\end{algorithm}

\subsection{Additional Profit Allocation}
\subsubsection{Shapley Value}
Energy sectors are playing a cooperative game to reduce the total operational cost. Thus, we design a Shapley value-based incentive mechanism for additional profit allocation.  
Shapley value has been widely adopted to measure the members' contributions to the collaboration earning \cite{fang2020improved} because of various game theoretical properties \cite{chau2022rkhs}. However, the allocation of Shapley value may be negative, which makes some sectors reluctant to join the end-to-end model \cite{pinson2022regression}. The zero-Shapley value \cite{liu2020absolute} is considered in this work and the profit allocated to section nn is:
\begin{equation}\label{shap_eq}
    v_{n}=\frac{1}{|\mathcal{N}|} \sum_{S \subseteq \mathcal{N} \backslash\left\{n\right\}} \frac{1}{\left(\begin{array}{c}
|\mathcal{N}|-1 \\
|S|
\end{array}\right)}[V\left(S \cup\left\{n\right\}\right)-V(S)]^{+}
\end{equation}
where $V(S)$ is the value of the cooperation formed by sectors union $S$, $V\left(S \cup\left\{n\right\}\right)$ is the value of the cooperation formed by sectors union $S \cup\left\{n\right\}$, $[\cdot]^{+}=\max \{0,\cdot\}$.

However, the zero-shapley value does not satisfy the budget balance property: the sum of revenues is equal to the sum of payments. So we need to adopt a normalized function $\Gamma$ to allocate the profits according to $v_n,n\in \mathcal{N}$.
\begin{equation}
    \Gamma(v_n)=\frac{v_n}{\sum_{i\in \mathcal{N}}v_i}(V(\mathcal{N})-V({\varnothing})) 
\end{equation}

\begin{figure}[t]
    \centering
\includegraphics[scale=0.48]{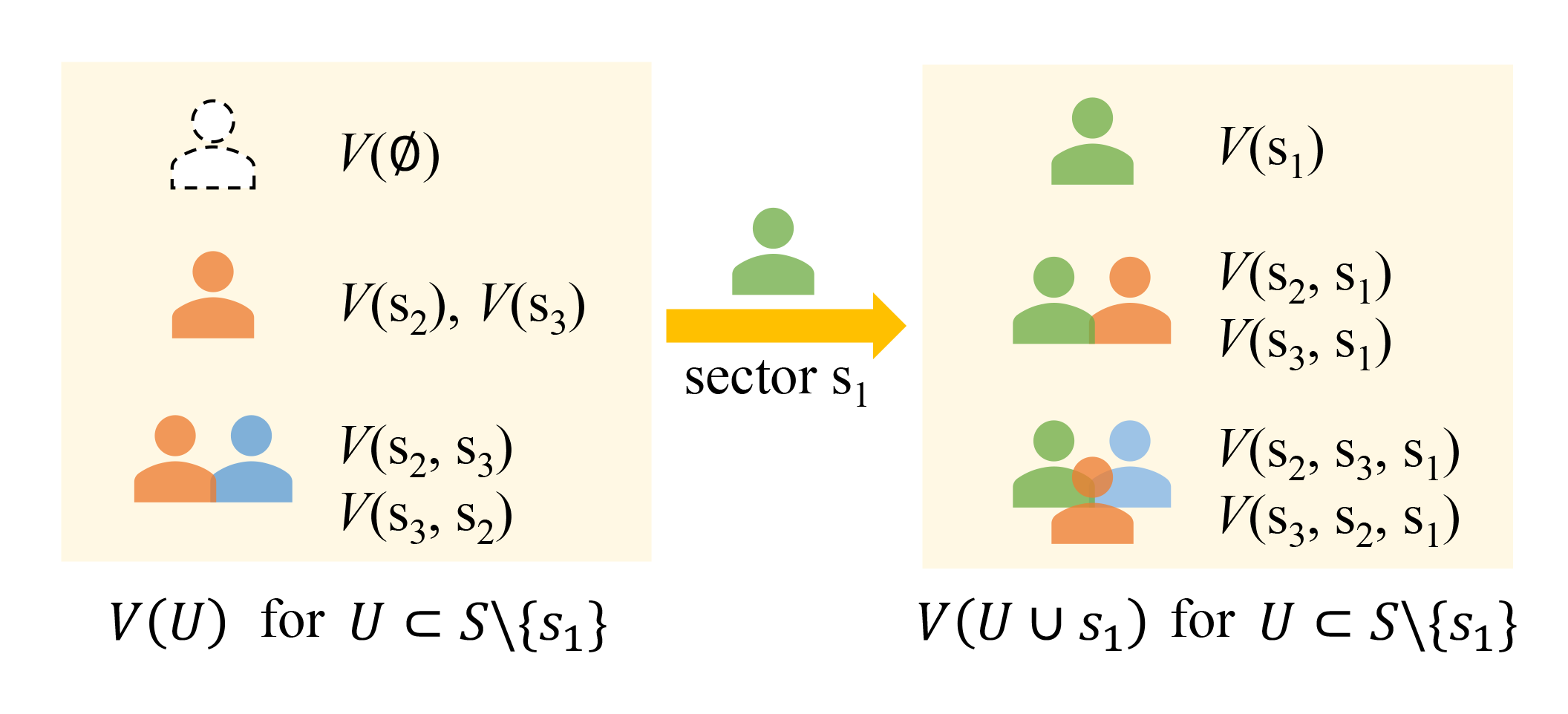}
    \caption{Profits allocation strategy with three sectors.}
    \label{shap}
\end{figure}

Fig. \ref{shap} illustrates the basic idea of Shapley value-based profit allocation with three sectors in MES.  The profits allocated to sector $n$ are measured by calculating the difference in profit before and after the sector participates in cooperation with different sector combinations \cite{dabbagh2015risk}. Therefore, we need to consider the situation where some sectors within the MES do not participate in the end-to-end modeling, which is referred to as ``partially integrated" cooperation. %The next subsection will discuss how to quantify additional profits earned by a ``partially integrated" cooperation.

\subsubsection{Value of ``Partially Integrated" Cooperation} This subsection first describes the scenario of ``partially integrated" cooperation and then quantifies its value.

When only the sectors in $U$ cooperate with the MES operator and sectors, the sectors in $\mathcal{N}\backslash U$ will remain their model parameters unchanged (denoted as $\overline{w}_n$) and only submit final forecasts $M_n(X_n;\overline{w}_n)|_{n \in \mathcal{N}\backslash U}$ to the operator. This scenario is described as:
\begin{equation}\label{cnu}
\mathop{\min}\limits_{z;w_n|_{n \in U}}C(z,M_n(X_n,w_n)|_{n \in U},M_n(X_n;\overline{w}_n)|_{n \in \mathcal{N}\backslash U})
\end{equation}
where $U$ denotes the set of sectors that cooperate with the operator; $C_{U}$ denotes the operation costs of the ``partially integrated" end-to-end model. When $U$ equals $\varnothing$, this scenario degrades to the traditional FTO approach.

Algorithm \ref{alg2} can also be applied to scenarios where only parts of sectors are willing to cooperate with the MES operator. After the basic model development procedure, the parameters of models $\overline{w}_n$ belonging to $U$ are fixed. At the end-to-end modeling steps, only the sectors in $U$ update their model parameters $w_n|_{n\in U}$ with the gradient given by the operator (lines 17 to 20). Once the model updating of sectors in $U$ is completed, the operation costs $C_{U}$ will be calculated with \eqref{cnu}.
The additional profits derived by the ``partially integrated" end-to-end model are regarded as the value of the cooperation formed by sectors in $U$:
\begin{equation}\label{part cooperation valuation}
V(U)=C_{U}-C_{\varnothing}
\end{equation}
where $C_U$ is the operation costs if only sectors in $U$ cooperate with the MES operator.

\section{Case Studies}
\label{CS}
This section conducts comprehensive experiments to examine the end-to-end model and showcase how to quantify the value of the multi-energy load data.

\subsection{Experimental Setups}
\begin{figure}[t]
\centering
\includegraphics[scale=0.48]
    {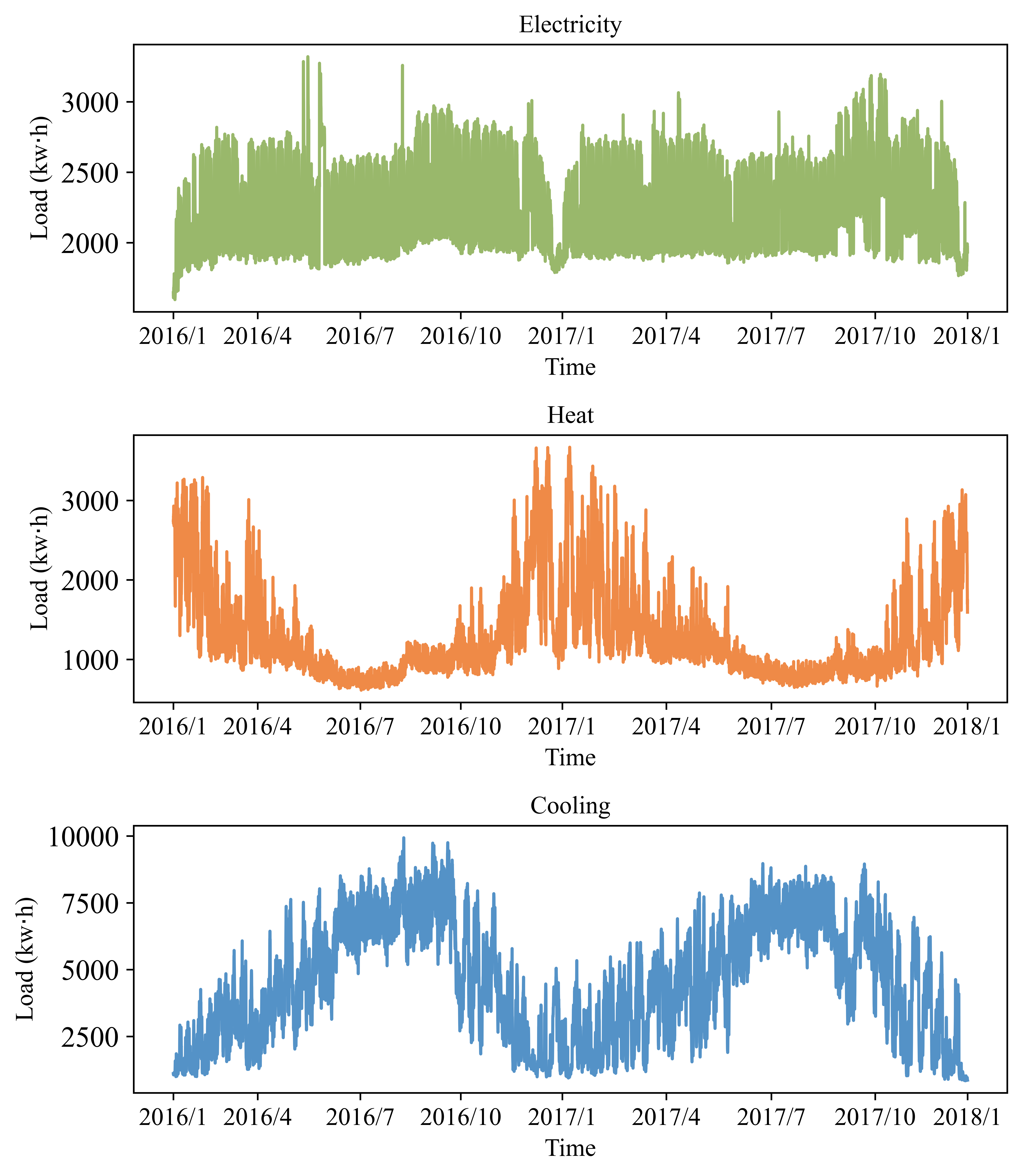}
    \caption{Multi-energy load profiles in MES in 2016 and 2017.}% The yellow arrow represents the steps that can be conducted easily, while the red arrow indicates the main challenge of end-to-end modeling.}
    \label{load}
\end{figure}

The case studies are conducted on a public dataset called Building Data Genome Project 2 \cite{miller2020building}.  Specifically, the Austin station dataset is employed, which includes hourly load data for electricity, heat, and cooling collected at the University of Texas. The multi-energy loads are shown in Fig. \ref{load}. The load data in 2016 and 2017 are used as the training and testing datasets, respectively. 

\begin{figure}[t]
\includegraphics[scale=0.4]
    {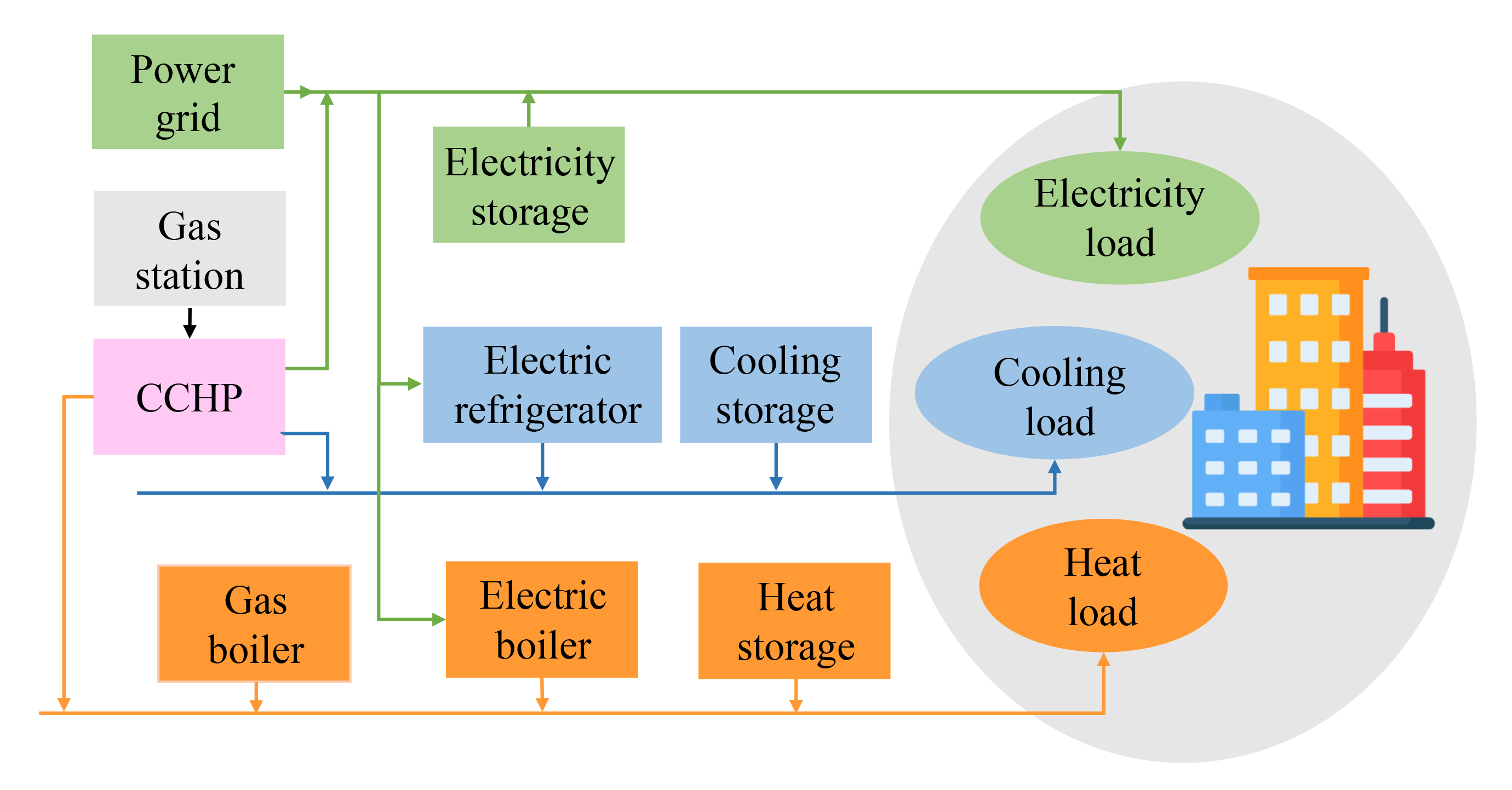}
    \caption{The structure of the studied MES.}% The yellow arrow represents the steps that can be conducted easily, while the red arrow indicates the main challenge of end-to-end modeling.}
    \label{system}
\end{figure}

Fig. \ref{system} shows the studied MES, including cooling, heat, and power (CCHP), gas boiler, electric boiler, electric refrigerator, and different storages. The turbine used in the studied MES is a backpressure turbine, which means that the generated electricity and heat are in a certain proportion.

The MES operation includes day-ahead scheduling and intra-day adjustment. The operator first schedules the energy converters based on multi-energy load forecasts in the day ahead. For the electricity sector, the electricity purchased from the power grid and generated from CCHP will be utilized to meet the electricity demands of residential areas as well as power electric boilers and refrigerators (green line in Fig. \ref{system}). For the heat sector, the heat produced by gas boilers, electric boilers, and CCHP will be utilized to meet the heat demand (orange line). The cooling sector will utilize both CCHP and electric refrigerators to meet the cooling load (blue line). During daily operations, due to the potential gap between actual load and load forecast, operators have three ways to maintain supply-demand balance: (1) purchase electricity from the grid temporarily, (2) adjust the output of energy converters within backup capacity constraints, and (3) utilize of energy storage for energy charging and discharging.

\subsection{End-to-End Modeling Results}
This subsection depicts the end-to-end modeling results, including forecasting performance and operation costs.

Mean absolute error (MAE), root mean square error (RMSE), and mean average percent error (MAPE) are used to evaluate the forecasting performance:
\begin{equation}
    \text{MAE}=\frac{1}{|D|}\sum_{i\in D}|{M_i}-M_i^{\text{real}}|\end{equation}
\begin{equation}\text{RMSE}=\sqrt{\frac{1}{|D|}\sum_{i\in D}({M_i}-M_i^{\text{real}})^{2}}\end{equation}
\begin{equation}\text{MAPE}=\frac{100\%}{|D|}\sum_{i\in D}|\frac{{M_i}-M_i^{\text{real}}}{M_i^{\text{real}}}|
\end{equation}
where $\hat{M_i}$, $M_i^{\text{real}}$, and $D$ denote the $i^{\text{th}}$ forecasts, the $i^{\text{th}}$ actual loads, the test data set, respectively. 

The load forecasting performance of the three sectors is given in Table \ref{table_accuracy_variation}, where the benchmark refers to the forecasting models that are separately trained with MSE. The accuracy of the electricity and cooling sector measured by MAE, RMSE, and MAPE are similar to the benchmarks. The accuracy of heat load prediction has experienced a slight decrease.

In the absence of a significant impact on accuracy, we will focus on MES operation costs. Table \ref{table_monthly_cost} shows the average daily operation costs in 2017. FTO refers to the average daily operation costs when $M$ is given by the models trained with MSE separately; Ideal corresponds to the operation costs when the day-ahead forecasts for $M$ are perfectly accurate. The operation costs during the summer are higher compared to other seasons due to the smaller heat load and higher cooling load. In this situation, the output of CCHP is determined by the heat load, which requires a significant amount of electricity to meet the cooling demand, leading to increased operation costs.

\begin{figure}[t]
    \includegraphics[scale=0.45]{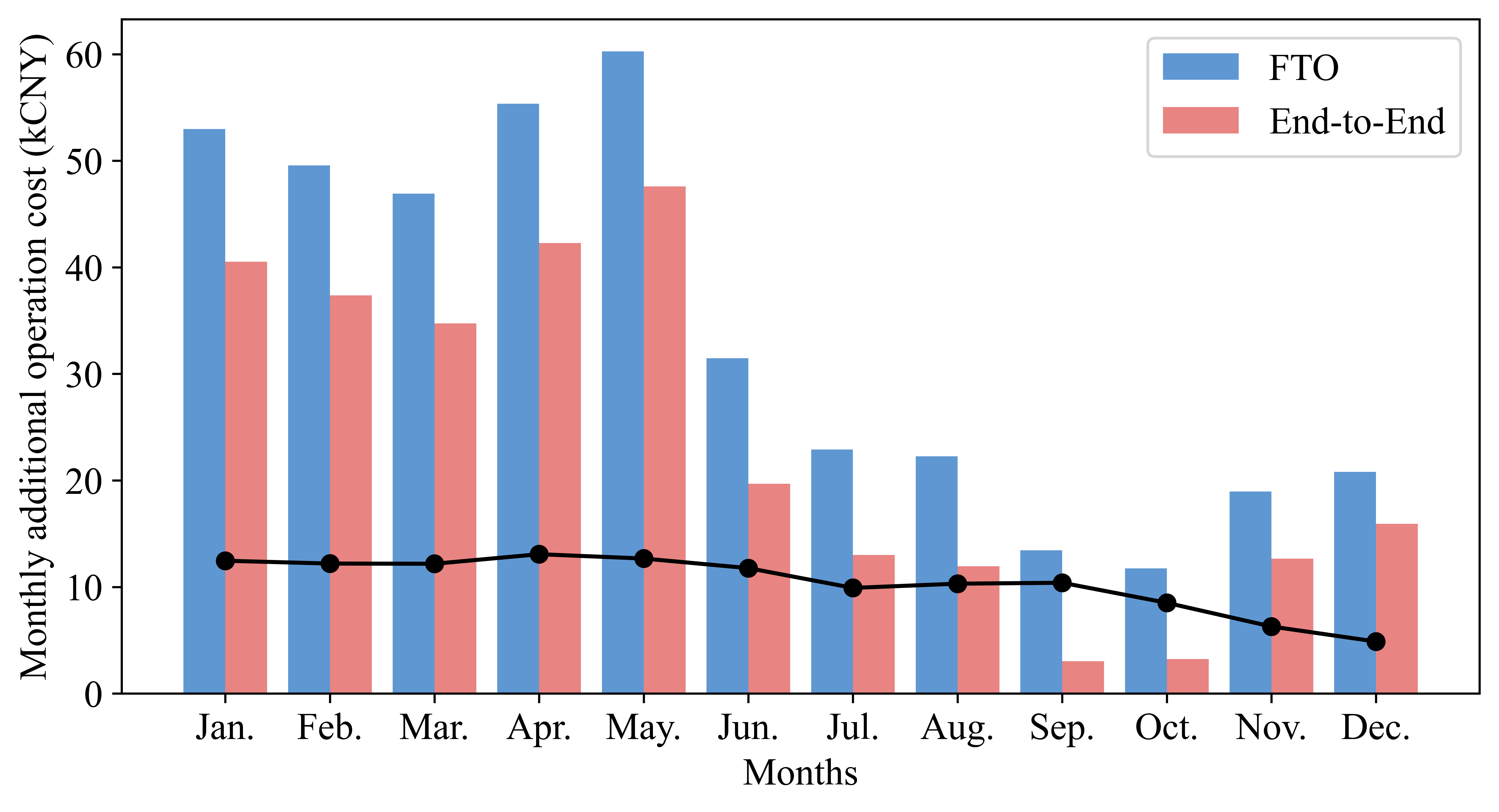}
    \caption{Monthly additional operation cost of FTO and end-to-end model compared to ideal cost (kCNY)}
    \label{monthly cost}
\end{figure}

The additional operation costs of FTO and end-to-end modeling compared to ideal costs are illustrated in Fig. \ref{monthly cost}. The black line in the graph represents the monthly operation costs savings compared to FTO achieved by end-to-end modeling. The average daily costs of the end-to-end modeling for each month in 2017 significantly decreased compared to FTO. The additional operation cost of the end-to-end model has been significantly reduced, especially in Sep. and Oct.  The annual total operation costs in the ideal case are 31012.06 kCNY. The values for the end-to-end model and FTO are 31294.04 kCNY and 31418.71 kCNY, respectively. Compared with the ideal operation costs, the forecasting errors result in 1.31\% additional operation cost when models are trained with MSE. In comparison, the forecasting errors just lead to 0.91\% additional operation cost when all sectors are enrolled in end-to-end modeling. This indicates that the operation achieves a 0.40\% reduction, resulting in annual cost savings of 124.66 kCNY. These findings highlight the significance of indirect data sharing between sectors.
\begin{table}[t]
\centering
\caption{Load forecasting performance of three sectors in MES}
\label{table_accuracy_variation}
\begin{tabular}{@{}ccccc@{}}
\toprule
                      &        \textbf{Model}      & \textbf{MAE}    & \textbf{RMSE}   &   \textbf{MAPE}   \\ \midrule
\multirow{3}{*}{Electricity
sector} & Benchmark     & 82.592 & 113.102 & 3.565 \\
                      & End-to-End & 82.383 & 112.951  & 3.562 \\
                      & Accuracy Variation          & 0.25\% & 0.13\% & 0.00\% \\ \midrule
\multirow{3}{*}{Heat sector} &  Benchmark      & 120.967 & 178.530  & 9.074 \\
                      & End-to-End & 121.331 & 181.348  & 9.216 \\
                      & Accuracy Variation           & -0.30\% & -1.58\%  & -0.14 \%     \\ \midrule
\multirow{3}{*}{Cooling sector} &  Benchmark     & 487.406 & 626.781 & 12.895 \\
                      & End-to-End & 477.677 & 615.956 & 12.799 \\
                      & Accuracy Variation           & 2.00\% & 1.73\%  & 0.10\%      \\ \bottomrule
\end{tabular}
\end{table}

% Please add the following required packages to your document preamble:
% \usepackage{booktabs}
\begin{table}[t]
\centering
\caption{Daily operation costs for 12 months in 2017 (kCNY)}
\label{table_monthly_cost}
\begin{tabular}{@{}ccccc@{}}
\toprule
     & \textbf{FTO} & \textbf{End-to-End} & \textbf{Ideal} & \textbf{Improvement} \\ \midrule
Jan. & 88.517               & 88.115                        & 86.808              & 0.454  \%                  \\ 
Feb. & 87.500               & 87.064                        & 85.730               & 0.498  \%                  \\ 
Mar. & 86.687               & 86.294                        & 85.174               & 0.453  \%                  \\ 
Apr. & 88.015               & 87.579                        & 86.170               & 0.495  \%                  \\ 
May  & 88.282               & 87.873                        & 86.338               & 0.463  \%                  \\ 
Jun. & 92.834               & 92.442                        & 91.786               & 0.423  \%                  \\ 
Jul. & 95.141               & 94.821                        & 94.402               & 0.336  \%                 \\ 
Aug. & 93.729               & 93.373                        & 92.961               & 0.379  \%                  \\ 
Sep. & 92.823               & 92.477                        & 92.375               & 0.373  \%                  \\ 
Oct. & 87.576              & 87.301                        & 87.197               & 0.314  \%                 \\ 
Nov. & 86.505               & 86.231                       & 85.681               & 0.316  \%                  \\ 
Dec. & 85.328               & 85.140                        & 84.527               & 0.220  \%                  \\ \bottomrule
\end{tabular}
\end{table}

\begin{figure}[t]
    \centering
    \includegraphics[scale=0.55]{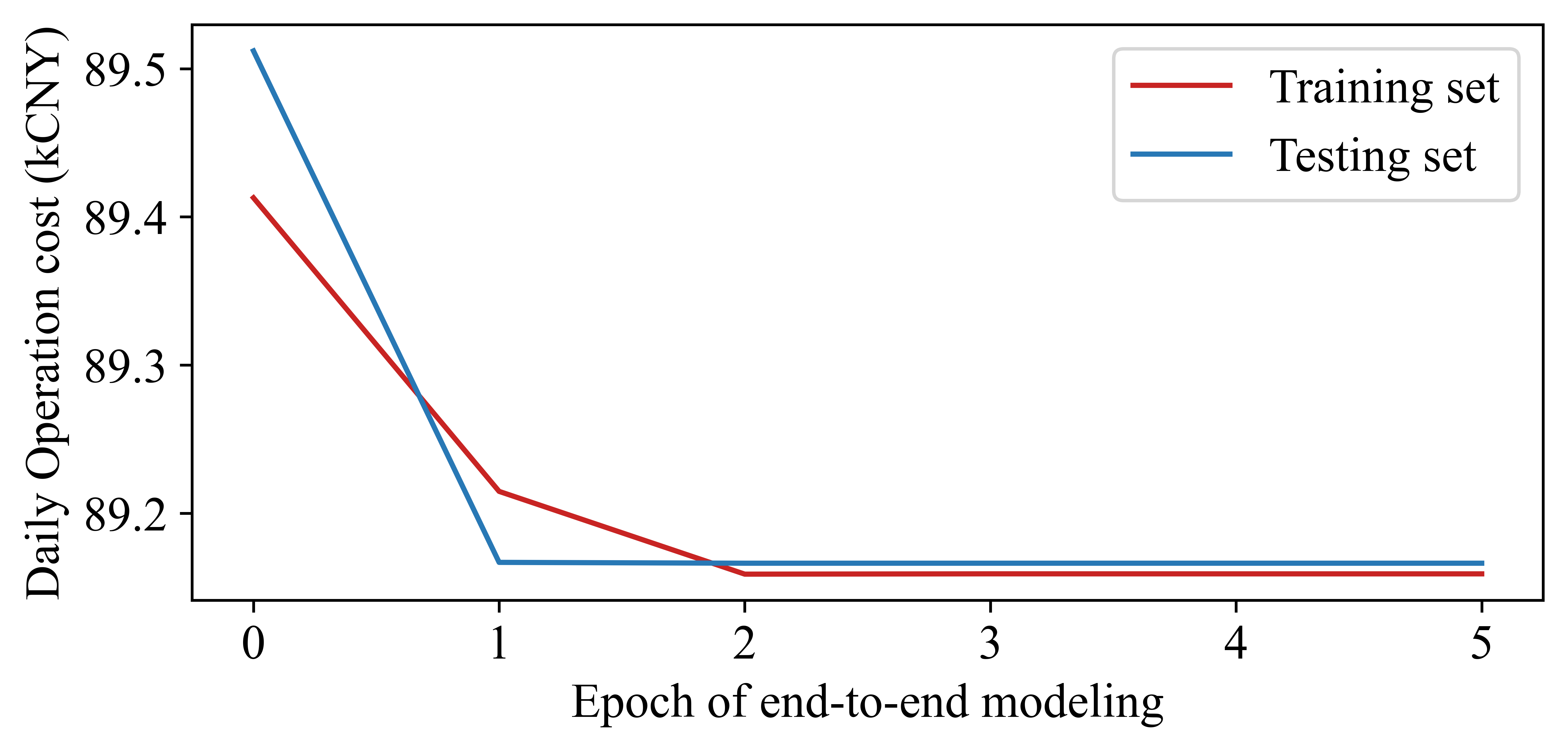}
    \caption{Daily average cost of the training/testing dataset in the end-to-end modeling process}
    \label{loss}
\end{figure}

The variation of the daily average costs of the training and testing dataset in the end-to-end modeling process is shown in Fig. \ref{loss}. It shows the model can reach convergence in just two epochs after the basic model development procedure, indicating that the proposed method possesses favorable convergence properties. 

\subsection{Data Valuation Results}
\begin{figure*}[t]
    \includegraphics[scale=0.5]{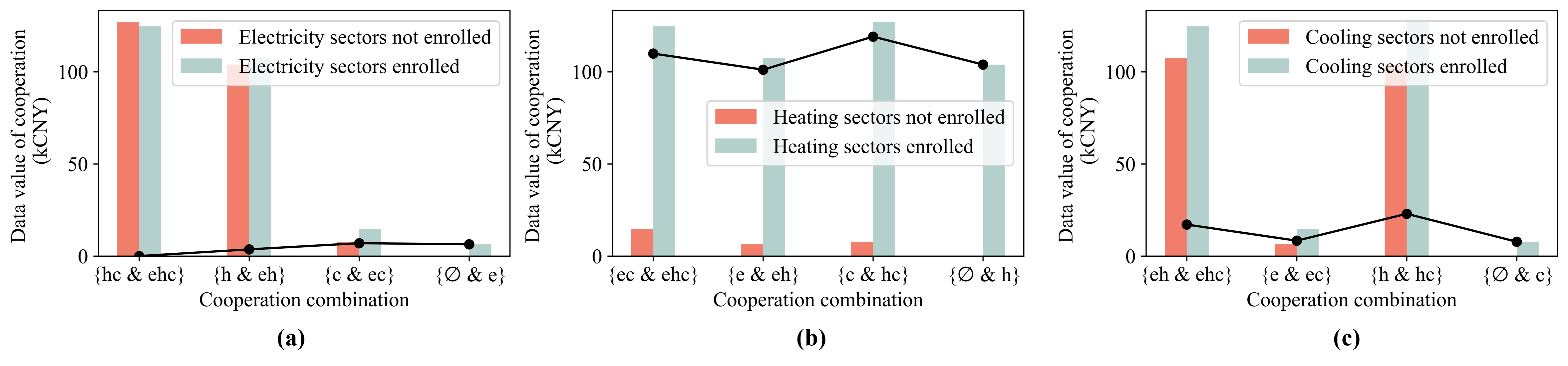}
    \caption{ 
    The data value of cooperation formed by different sectors combination. \textbf{(a)} With and without the involvement of the electricity sector. \textbf{(b)} With and without the involvement of the heat sector.  \textbf{(c)} With and without the involvement of the cooling sector.}
    \label{shap_fig}
\end{figure*}

\begin{figure}[t]
    \includegraphics[scale=0.45]{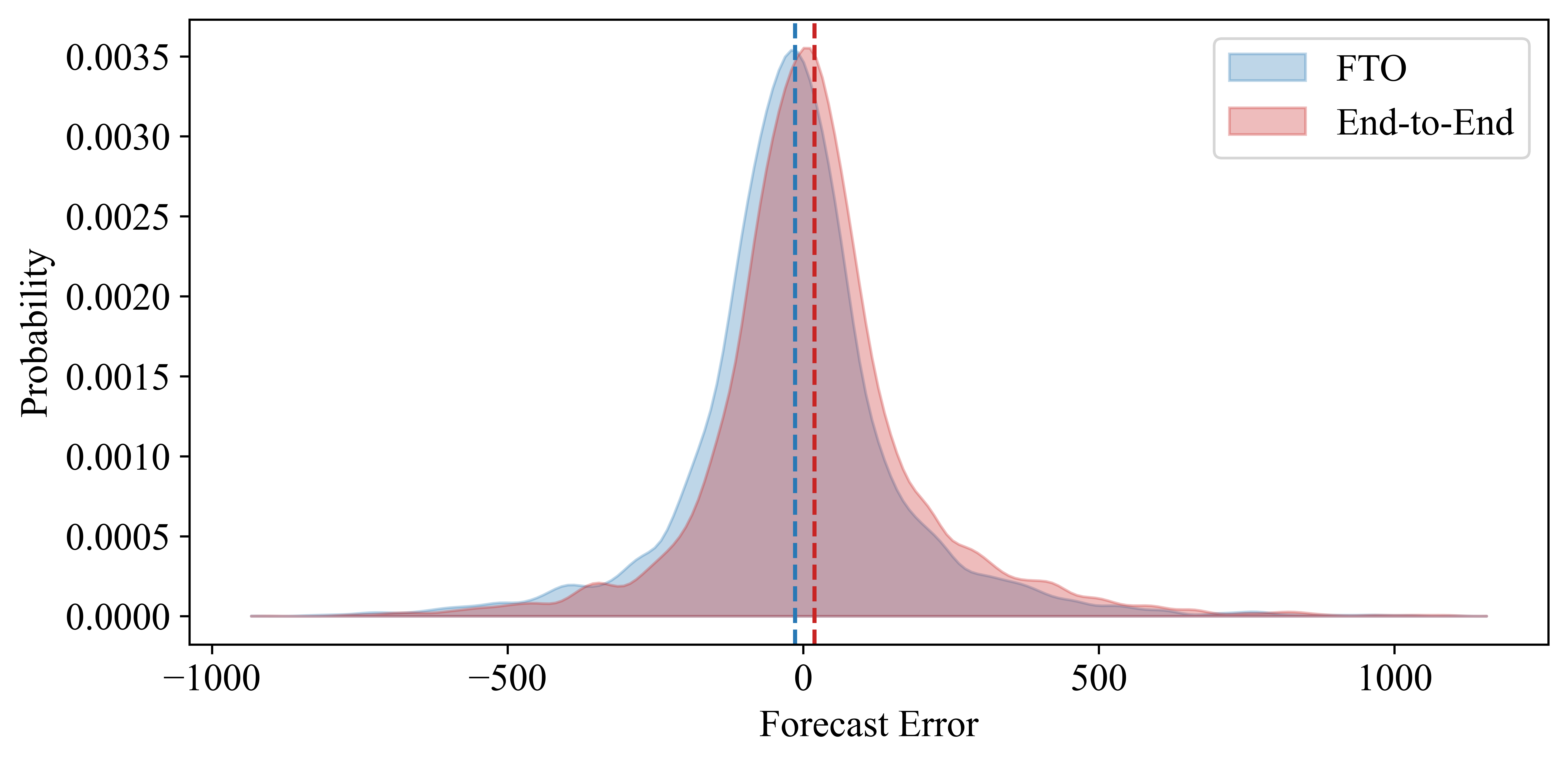}
    \caption{The distribution of the forecasts error of heat sector}
    \label{forecasts error}
\end{figure}

Fig. \ref{shap_fig} shows the additional profits from indirect data sharing in cooperation combination. The electricity, heat, and cooling sectors in cooperation are denoted as $\textit{e}$, $\textit{h}$, and $\textit{c}$ for the convenience of illustration. 
The red bar, green bar, and black point denote the $V(S)$, $V(S \cup\left\{n\right\})$, and $V\left(S \cup\left\{n\right\}\right)-V(S)$ in \eqref{shap_eq}, respectively. 
In (\textbf{a}), the enrollment of the electricity sector does not contribute significantly to additional profits. There are two possible reasons: (1) the forecasting accuracy of the electricity sector is relatively high, and the variation of forecasting accuracy within the reserve capacity in the electricity sector has little effect on operation costs. (2) The deviation of the electricity price in intra-day and day-ahead is relatively small, and the additional operation costs resulting from electricity forecasting errors are relatively small. In (\textbf{b}), it is evident that the earned additional profits when the heat sector cooperates with the MES operator can be markedly improved. When the heat load is relatively small, the output of the CCHP unit will be determined by the heat load. As shown in Fig. \ref{forecasts error}, the distribution of forecasting errors moves in the direction of over-forecasting. Over-forecasting the heat load prediction can effectively increase the CCHP output and thus achieve a significant cost reduction. In (\textbf{c}), the enrollments of the cooling sector can slightly decrease the operation costs, especially when the heat sector cooperates with the MES operator at the same time. 

%In Fig. ref{shap_fig}, the bar represents the additional profits that can be obtained through collaboration, which is also considered the data value of the sectors involved in the cooperation. 

Table \ref{profits allocation} lists the detailed profit allocation result and calculation method. The total data value of three sectors $V(e,h,c)$  is 124.66 kCNY ($C_{\{e,h,c\}}-C_{\varnothing}$). According to \eqref{shap_eq}, the data valuations of the electricity, heat, and cooling sectors are 3.89 kCNY, 107.35 kCNY, and 13.43 kCNY, respectively. 

\begin{table*}[t]
\centering
\caption{Profit Allocation result (kCNY)}
\label{profits allocation}
\begin{tabular}{|cc|cccccccc|}
\hline
\multicolumn{2}{|c|}{Combinations}                                                          & \multicolumn{1}{c|}{$e$, $h$, $c$} & \multicolumn{1}{c|}{$e$, $h$} & \multicolumn{1}{c|}{$e$, $c$} & \multicolumn{1}{c|}{$h$, $c$} & \multicolumn{1}{c|}{$e$}  & \multicolumn{1}{c|}{$h$}  & \multicolumn{1}{c|}{$c$}  & $\varnothing$    \\ \hline
\multicolumn{2}{|c|}{Operation   costs $C_{(\cdot)}$}                                                     & \multicolumn{1}{c|}{31294.04}    & \multicolumn{1}{c|}{31291.83}  & \multicolumn{1}{c|}{31311.15}  & \multicolumn{1}{c|}{31403.95}  & \multicolumn{1}{c|}{31412.30} & \multicolumn{1}{c|}{31314.79} & \multicolumn{1}{c|}{31410.94} & 31418.71\\ \hline
\multicolumn{2}{|c|}{Data valuation $V(\cdot)$}                                                            & \multicolumn{1}{c|}{124.66}      & \multicolumn{1}{c|}{126.87}    & \multicolumn{1}{c|}{107.56}    & \multicolumn{1}{c|}{14.76}    & \multicolumn{1}{c|}{6.40}   & \multicolumn{1}{c|}{103.92}   & \multicolumn{1}{c|}{7.77}   & 0      \\ \hline
\multicolumn{1}{|c|}{\multirow{6}{*}{Contributions}} & \multirow{2}{*}{Electricity sector} & \multicolumn{8}{c|}{\multirow{2}{*}{$\Gamma(\frac{[V(e,h,c)-V(h,c)]^{+}+\frac{1}{2}[V(e,h)-V(h)]^{+}+\frac{1}{2}[V(e,c)-V(c)]^{+}+[V(e)-V({\varnothing})]^{+}}{3})=3.89$}}                                                                                                                                                                                   \\
\multicolumn{1}{|c|}{}                               &                                      & \multicolumn{8}{c|}{\multirow{2}{*}{}}                                                                                                                                                                                                          \\ \cline{2-10} 
\multicolumn{1}{|c|}{}                               & \multirow{2}{*}{Heat sector}             & \multicolumn{8}{c|}{\multirow{2}{*}{$\Gamma(\frac{[V(e,h,c)-V(e,c)]^{+}+\frac{1}{2}[V(h,c)-V(c)]^{+}+\frac{1}{2}[V(e,h)-V(e)]^{+}+[V(h)-V({\varnothing})]^{+}}{3})=107.35$}}                                                                                                                                                                                   \\
\multicolumn{1}{|c|}{}                               &                                      & \multicolumn{8}{c|}{}                                                                                                                                                                                                          \\ \cline{2-10} 
\multicolumn{1}{|c|}{}                               & \multirow{2}{*}{Cooling sector}             & \multicolumn{8}{c|}{\multirow{2}{*}{$\Gamma(\frac{[V(e,h,c)-V(e,h)]^{+}+\frac{1}{2}[V(e,c)-V(e)]^{+}+\frac{1}{2}[V(h,c)-V(h)]^{+}+[V(c)-V({\varnothing})]^{+}}{3})=13.43$}}                                                                                                                                                                                    \\
\multicolumn{1}{|c|}{}                               &                                      & \multicolumn{8}{c|}{}                                                                                                                                                                                                          \\ \hline
\end{tabular}
\end{table*}

The valuation proposed in this paper exhibits significant advantages in terms of both allocation fairness and incentive effect. From the perspective of allocation fairness, the heat sector's enrollment significantly reduces the operation costs, as shown in Fig. \ref{shap_fig}, justifying the allocation of the largest share of profits to it. As for the electricity sector, it has made the least contribution, so it should share the least share of profits. From the perspective of intensive effect, the electricity and cooling sectors not only achieved improvements in model accuracy but also acquired modest profits through their participation in the end-to-end model. As for the heat sector, the degradation of the forecasting accuracy caused by enrolling the end-to-end model is compensated with a reasonable economic reward. 

%\begin{figure}[t]
  %  \includegraphics[scale=0.5]{./monthly allocation.png}
  %  \caption{Monthly additional profits allocation among sectors.}
   % \label{monthly_allocation}
%\end{figure} 

\section{Conclusions and Future Works}
\label{conclusion}
This paper proposes an end-to-end approach for quantifying the additional profits of the multi-energy load data owned by different sectors in MES and then designs an incentive mechanism to allocate additional profits from data sharing to different sectors in MES according to their contributions.

The experiments have been conducted with the data collected from the real world. The results indicate that the proposed end-to-end modeling approach can achieve the operation cost reduction under the premise of not affecting the forecasting accuracy of various sectors in MES.
%The results show that the forecasting accuracy of the electricity and cooling sectors improved while the accuracy of the heat sector degraded. Compared to the traditional FTO approach, The operation costs decrease significantly after end-to-end modeling. Such operation costs reductions are derived through indirect data sharing between the end-to-end model.
The data valuation results illustrate that the proposed valuation approach can thoroughly uncover the value of data through indirect data sharing among sectors in MES. This approach offers a direct economic measurement rather than analyzing data utility.
In the profit allocation, the contribution of each sector to the total operation cost-saving has been equitably measured. Our profit allocation strategy can effectively incentivize sectors to cooperate with the MES operator to achieve better operation costs because all sectors can benefit from the cooperation in accuracy improvement or economic compensation. 
%Through the visualization of the relationships between MES operation costs and load forecasting errors, this study identifies two factors that elucidate why cost savings can still be attained, even though some sectors experiencing a decline in forecasting accuracy. 

Although our current valuation framework is effective, it relies on a trusted third party to calculate the gradient in the end-to-end model training process and assumes that all sectors involved are honest. In the future, we will develop a privacy-preserving data valuation framework and explore new design mechanisms that encourage all sectors to be truthful and honest.

\bibliographystyle{IEEEtran}
 \bibliographystyle{elsarticle-num} 
 %%\bibliography{cas-refs}

\end{document}